\documentclass[12pt,manuscript]{emulateapj}
\usepackage{slashbox}

\shorttitle{Kernel-phase Fizeau Interferometry}
\shortauthors{F. Martinache}

\begin{document}

\bibliographystyle{apj}

\title{Kernel-phase in Fizeau Interferometry}

\author{Frantz Martinache\altaffilmark{1}}

\altaffiltext{1}{Subaru Telescope, Hilo HI}

\begin{abstract}
The detection of high contrast companions at small angular separation
appears feasible in conventional direct images using the
self-calibration properties of interferometric observable quantities.
The friendly notion of closure-phase, which is key to the recent
observational successes of non-redundant aperture masking
interferometry used with Adaptive Optics, appears to be one example of
a wide family of observable quantities that are not contaminated by
phase-noise.
In the high-Strehl regime, soon to be available thanks to the coming
generation of extreme Adaptive Optics systems on ground based
telescopes, and already available from space, closure-phase like
information can be extracted from any direct image, even taken with 
a redundant aperture.
These new phase-noise immune observable quantities, called
kernel-phases, are determined a-priori from the knowledge of the
geometry of the pupil only.
Re-analysis of archive data acquired with the Hubble Space Telescope
NICMOS instrument, using this new kernel-phase algorithm demonstrates
the power of the method as it clearly detects and locates with
milli-arcsecond precision a known companion to a star at angular
separation less than the diffraction limit.
\end{abstract}


\section{Phase in the Fourier Plane}

Only two parameters essentially determine whether a source is
detectable during an observation: its brightness at the wavelength
$\lambda$ of interest and the angular resolution necessary to separate
the source or feature from its direct environment.
The angular resolution is ultimately constrained by the diffraction of
the telescope, and astronomers usually follow the rule of thumb known
as the Rayleigh criterion, stating that to be resolved, two sources
need to be separated by $1.22 \lambda/D$, where $D$ is the diameter of
the telescope used, to design their observations.

The development of optical interferometry has however made this
criterion obsolete: thanks to the exquisite level of calibration it
permits, interferometry indeed makes it possible to detect sources or
constraint the extent of features around objects at separations
significantly smaller than the diffraction limit.
Even at the scale of one single telescope, the results obtained with
the technique known as non-redundant masking (NRM) interferometry,
first, seeing-limited \citep{1987Natur.328..694H, 1988AJ.....95.1278R}
and more recently used with Adaptive Optics (AO) systems
\citep{2000PASP..112..555T, 2006SPIE.6272E.103T, 2006ApJ...650L.131L,
  2008ApJ...678..463I, 2008ApJ...679..762K, 2009ApJ...695.1183M}
demonstrate the relevance of this technique for the detection of
structures at small angular separation, that would not be accessible
from conventional AO images \citep{2004SPIE.5491.1120R}.

Even if it only uses one single telescope, in order to reach this
level of resolution, one however needs to accept that the familiar
product called ``image'' may not necessarily constitute the best final
data-product.
Instead, when interested in high-angular resolution properties of
partially resolved objects, it is convenient to derive information not
from the image itself, but from its Fourier-transform
counterpart. This information, known as complex visibility, is
extracted from the Fourier-plane, calibrated and then tested against a
model of the observed object.

In optical interferometry, this approach is often mandatory: the
paucity of apertures ($N \sim 2-5$) and baselines make the content of
a direct (Fizeau) image of limited value.
Information rich images can be reconstructed after extraction of the
complex visibility function from the $u,v$ plane, but only with a
large ($N > 10$) number of apertures like in radio-interferometry, or
after using image synthesis.
The optical image reconstruction known as pupil densification that is
used in hypertelescopes \citep{1996A&AS..118..517L} does provide an
alternative, but again, only becomes compelling if a large number of
apertures is used \citep{2008SPIE.7013E.107L}.
But even when an image can be reconstructed from optical
interferometry measurements, e.g. the images of the binary Capella by
\citet{1996A&A...306L..13B}, the intensity map of the surface of
Altair by \citet{2007Sci...317..342M} or the spectacular images of the
disk eclipsing $\epsilon$ Aurigae \citep{2010Natur.464..870K},
quantitative characteristics of the sources can only be deduced from
the fit of the interferometric data by parametric models.
In the case of a marginally resolved binary star, precise measurements
of angular separation, orientation and contrast, with confidence
intervals, deduced from a model-fit of complex visibilities carry much
more scientific value than an image of ``blurry blobs''.

Visibilities in the Fourier-plane are complex numbers, whose amplitude
and phase are usually considered separately. This paper focuses on the
treatment of the phase and ignores the amplitude.
In general, the power contained at given spatial frequency is the
result of the coherent sum of $R$ random phasors, with $R$ a scalar
coding the redundancy of the spatial frequency.
In the presence of residual optical path differences (OPD), this
coherent sum of $R$ random phasors loses the phase information and
results in the formation of speckles in seeing-limited images with a
visible/IR telescope.
NRM-interferometry solves this problem, by discarding light with a
pupil mask designed so that each baseline is unique ($R=1$), which
makes the extraction of the phase possible. 

The phases alone, being corrupted by residual OPDs, are of restricted
interest. It is however possible to combine them to form what is known
as closure-phase \citep{1958MNRAS.118..276J}, that is the sum of three
phases measured by baselines forming a closed triangle. This
remarkable interferometric quantity (cf. the introduction to closure
phase by \citet{2000plbs.conf..203M}) exhibits a compelling property:
it rejects all residual pupil-plane phase errors. Moreover, because it
is determined from the analysis of the final science detector and not
on a separate arm wavefront sensor, it is also immune to non-common
path errors between the wavefront sensor and the science camera.

Once extracted and calibrated, the closure-phases can then be compared
to a parametric model, for instance of a binary star, to confirm or
infirm the presence of a companion around a given source, while
uncertainties provide contrast detection (i.e. sensitivity) limits.
This approach was successfully used by \citep{2006ApJ...650L.131L,
  2007ApJ...661..496M, 2008ApJ...678..463I, 2008ApJ...679..762K,
  2009ApJ...695.1183M}, who typically report sensitivity of 5-6
magnitudes in the near infrared at separations ranging from 0.5 to 4
$\lambda/D$.

This paper aims at generalizing the notion of closure-phase, and shows
that closure-phase like quantities, i.e. sharing the same property of
independence to pupil-plane phase errors, can be constructed even in
the case of redundant apertures.

\section{Kernel-phase}
\subsection{Linear model}
\label{sec:ker}

Whether contiguous (i.e. single-dish) or not (i.e. interferometric),
the 2D pupil of an imaging system can be discretized into a finite
collection of $N$ elementary sub-apertures.
One of these elementary sub-apertures taken as zero-phase reference,
the pupil-plane phase of a coherent point-like light source can be
written as a $N-1$-component vector $\varphi$.
Given that the image, or interferogram, of this source is sufficiently
sampled, then in the Fourier plane (a.k.a. $(u,v)$-plane in
interferometry) one will be able to sample up to $M$ phases, where $M$
is a function of the pupil geometry only. For a non-redundant array
made of $N$ elementary sub-apertures, the number of sampled $(u,v)$
phases is maximum $M = (^N_2)$. The same number of sub-apertures
organized in a redundant array, for instance following a regular grid,
produces significantly less distinct $(u,v)$ sample points as each
point receives the contribution of several pairs of sub-apertures. 

In most cases, since each point receives the sum of several random
phasors, both phase and amplitude are lost and cannot be simply
retrieved: this results in the formation of speckles.
However, if the Strehl is high enough, the complex amplitude
associated to the instrumental phase in one point of the pupil,
$\varphi_k$, can be approximated by $e^{i\varphi_k} \approx 1 +
i\varphi_k$.
Direct application of the approach is therefore for now, restricted to
space-borne diffraction-limited optical and mid-IR telescopes like HST
(cf. Section \ref{sec:hst}), but should also prove relevant to the
upcoming generation of extreme AO systems.

\begin{figure}
\plotone{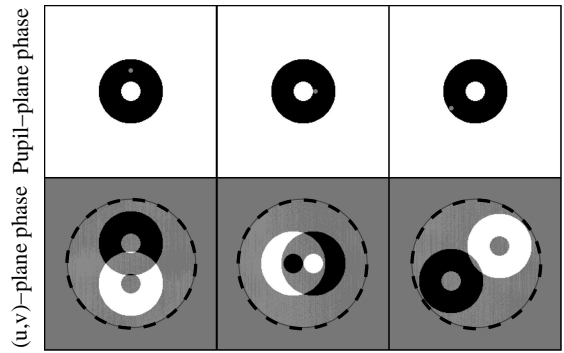}
\caption{
Iterative process for the determination of the transfer matrix
$\mathbf{A}$. The top row shows the sub-aperture of the full 2D pupil
(circular aperture with 30\% central obscuration) where a phase offset
is applied (three cases are represented). The bottom row shows the
resulting distribution of phase in the Fourier plane. The overlaid
dashed-line circle in the bottom row marks the cutoff spatial
frequency of the transfer function.
}
\label{fig:push}
\end{figure}

Given that the proposed approximation holds, while observing a point
source, the unknown (instrumental) phase distribution in the pupil
$\varphi$ can be related to the phases $\Phi$ measured in the Fourier
plane with a single linear operator.
To build an intuitive understanding of this relation, let us consider
the following scenarios:

\begin{itemize}
\item If the phase is constant across the entire pupil, then none of
  the baselines formed by any pair of elementary sub-apertures does
  record a phase difference, and the phase in the Fourier plane is
  zero everywhere.
\item If a phase offset $\delta_\phi$ is added to one single
  sub-aperture, then each baseline involving this aperture records a
  phase difference, which is exactly $\pm \delta_\phi$.
  Figure~\ref{fig:push} represents several such scenarios.
\item If the pupil-plane phase vector $\varphi$ is completely random,
  each of the $M$ samples in the Fourier-plane is then the average of
  $R$ phase differences on the pupil, where $R$ is the redundancy of
  the considered baseline.
\end{itemize}

To reproduce this behavior, the following linear model will be used:

\begin{equation}
\Phi = \mathbf{R^{-1}} \cdot \mathbf{A} \cdot \varphi,
\label{eq:transfer}
\end{equation}

\noindent
where $\Phi$ represents the $M$-component Fourier plane phase vector,
$\mathbf{R}$ a $M \times M$ diagonal matrix whose diagonal elements
code the redundancy of the baselines,
and $\mathbf{A}$ represents a $M \times (N-1)$ transfer matrix, whose
properties form the core of the discussion of this work. To be
complete, the model should also include the phase information
intrinsic to the observed source, represented by the term $\Phi_O$
that simply adds on top of the instrumental phase. 
One can then multiply both sides of the equation by the matrix
$\mathbf{R}$ so that it becomes:

\begin{equation}
\mathbf{R} \cdot \Phi = \mathbf{A} \cdot \varphi + \mathbf{R} \cdot \Phi_O.
\label{eq:all}
\end{equation}

While $\mathbf{R}$ and $\mathbf{A}$ could have been merged into one
single operator, there are intentionally kept distinct. The rationale
for this choice so is that the left hand side of Eq. \ref{eq:all},
i.e. the measurements, can be acquired by reading directly the
imaginary part of the complex visibility.
Given that the next (quadratic) term in the Taylor expansion of
$e^{i\varphi}$ being real, this makes the approximation valid to the
third order in phase. This also makes $\mathbf{A}$ of striking aspect
as it is then exclusively filled with values 0, 1 or -1.

If the matrix $\mathbf{A}$ were invertible, then the analysis of one
unique focal plane image of a single star (case corresponding to
Eq. \ref{eq:transfer})
would be sufficient to determine the instrumental phase $\varphi$ as
seen from the detector, and drive an AO system and/or delay lines.
Except for the special case of a non-redundant aperture, the problem
is however known to be degenerate, despite the larger number of
measures than unknowns ($M > N-1$).\footnote{The use of this model for
wavefront sensing purposes will be the object of another paper.}

As demonstrated by the successes of NRM-interferometry, a complete
characterization of the wavefront is not essential if one can
determine observable quantities that are pupil-phase independent.
The closure relations used in interferometry can be related to the
operator $\mathbf{A}$: these relations are simply linear combinations
(modelized by an operator $\mathbf{K}$) of rows of $\mathbf{A}$ that
produce 0, forming something known as the left null space of A:

\begin{equation}
  \mathbf{K} \cdot \mathbf{A} = \mathbf{0}.
  \label{eq:leftnull}
\end{equation}

For a non-redundant array, each closure relation will fill a row of
$\mathbf{K}$ with mostly zeroes, except in three positions
corresponding to the baselines forming a closing triangle, that will
contain 1 or -1.
These relations are however not the only possible ones, and less
trivial combinations, involving more than three rows at a time, can be
constructed. The total number of independent relations however remains
unchanged and is only imposed by the geometry of the array.

Although not impossible, finding the operator $\mathbf{K}$ ``by hand''
(i.e. finding a basis for the left null-space of $\mathbf{A}$) for a
redundant aperture is a tedious task, as the matrix $\mathbf{A}$ can
get quite large.
A very efficient way to do this is to calcutate the singular value
decomposition (SVD) of $\mathbf{A}^T$.
The SVD algorithm \citep{2002nreb.book.....V}, allows to decompose the
now $(N-1) \times M$ matrix $\mathbf{A}^T$ as the product of a $(N-1)
\times M$ column-orthogonal matrix $\mathbf{U}$, a $M \times M$
diagonal matrix $\mathbf{W}$ with positive or zero elements (the
so-called singular values) and the transpose of an $M \times M$
orthogonal matrix $\mathbf{V}$:

\begin{equation}
  \mathbf{A}^T = \mathbf{U} \cdot \mathbf{W} \cdot \mathbf{V}^T.
  \label{eq:svd}
\end{equation}

One relevant property of the SVD is that it explicitly constructs
orthonormal bases for both the null-space and the range of the matrix
$\mathbf{A}^T$.
Of particular interest here, are the columns of $\mathbf{V}$ that
correspond to singular values equal to zero: these vectors form an
orthonormal base for the null-space, also refered to as Kernel of
$\mathbf{A}^T$, that is exactly what is needed to fill in the rows of
$\mathbf{K}$.

If the observed target is not perfectly symmetric, and exhibits actual
phase information (i.e. $\Phi_O \neq 0$, see for instance
\citet{2000plbs.conf..203M}), Eq. \ref{eq:all} is required. 
Multiplying it with the left side operator $\mathbf{K}$ leads to a
new series of new phase-like quantities that are not contaminated by
instrumental phase, generalizing the notion of closure-phase
\citep{1986Natur.320..595B} on which NRM-interferometry from the
ground \citep{2000PASP..112..555T} and from space
\citep{2009SPIE.7440E..30S} rely entirely.

While not as immediately tangible as the notion of closure-phase, this
proposed generalization, hereafter refered to as kernel-phase (or
Ker-phase) since it relates to the kernel of the matrix
$\mathbf{A}^T$, exhibits a unique advantage over the classical
closure-phase: it is not restricted to non-redundant apertures and
makes it possible to extract phase-residual immune information from
images acquired from telescopes of arbitrary pupil geometry.

\begin{figure}
\plotone{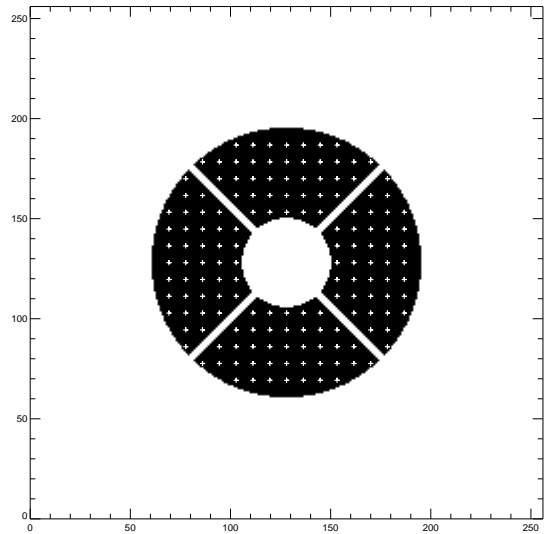}
\caption{
  Model for the geometry of the HST/NICMOS pupil and location of the
  sample points for the determination of the Ker-phase relations.
  The 156 sample points of the pupil fall on a regular square grid
  with a step of 1/16$^{th}$ of a pupil diameter that do not intersect
  with the central obstruction or the spider arms.
}
\label{fig:hst_pup}
\end{figure}

This has some obvious advantages over the restrictive non-redundant
case:

\begin{itemize}
\item throughput: non-redundant aperture masks exhibit a typical
  5-15\% throughput \citep{2007ApJ...661..496M}, and photon noise of
  the companion one tries to detect may be the dominant source of
  noise. Given that it benefits from the same phase-noise cancelling
  properties as closure-phase, for a given readout noise and exposure
  time, kernel-phase on an unmasked aperture offers an immediate boost
  in sensitivity (or dynamic range) on faint sources.
\item number of observable quantities: a common non-redundant aperture
  mask design exhibits nine sub-apertures, therefore forming
  $(^9_2)=36$ baselines and $(^8_2)=28$ independent closure phases
  \citep{2000plbs.conf..203M}. More independent kernel-phases can be
  extracted from the Fourier transform of a full-aperture image, which
  will provide a better characterization of the target.
\end{itemize}

Another incidental advantage is that, being a product of the SVD,
all the kernel-phase relations contained in $\mathbf{K}$ form an
orthonormal basis, and therefore do not introduce correlation in the
data.
A consequence is that manipulating Ker-phases does not require to
keep track of the covariance matrix used for closure-phases in masking
interferometry, which simplifies their interpretation.

\subsection{Calibration}

In discretizing the pupil into a finite number of sub-apertures, one
important assumption is made: the phase (or more generally, the
complex amplitude of the electric field) is assumed to be uniform
within each sub-aperture. Yet even for a space borne telescope, in the
absence of atmosphere, this is only an idealization as small scale
structures like polishing imperfections of the primary mirror for
instance, will impact, to some extent, the value of the Ker-phases.
This issue is not proper to Ker-phases and also affects
closure-phases. Thus, unless perfect (i.e. single-mode) spatial
filtering is performed within each sub-aperture of a non-redundant
array, the closure phase on a point source is not exactly be zero.

This effect can somewhat be mitigated by substracting from the
Ker-phases of a science target, the Ker-phase signal measured on a
point source observed in identical conditions. NRM-interferometry
results reported in \citet{2009ApJ...695.1183M} for instance, make
extensive use of this kind of calibration: from the ground, this
approach is very powerful as it makes it possible to calibrate other
sources of systematic errors like the effect of broadband filters
which smear out the Fourier plane and differential atmospheric
refraction.
From space, this may not be as essential depending on the science
goal: if the Ker-phases obtained on a binary system are
non-calibrated, then they will simply contain a systematic error term
that will limit the achievable contrast.

\section{Kernel phase analysis of HST/NICMOS data}
\label{sec:hst}

While the kernel-phase approach may prove difficult to apply to ground
based observations until extreme Adaptive Optics become available,
it can readily be applied to diffraction-limited observations made
from space.
It is tested here on a series of non-coronagraphic narrow band images
acquired with the Near-Infrared Camera and Multiobject Spectrometer
(NICMOS) onboard the Hubble Space Telescope. Two datasets acquired
with the NICMOS1 in the F190N filter on two distinct objects are
used: the first target is a calibration star, SAO 179809, which was
observed in 1998; the second is is the high-proper motion star GJ 164,
around which a companion was astrometrically discovered and whose
existence was confirmed after PSF modelization and substraction of
these NICMOS1 images by \citet{2004ApJ...617.1323P}. This latter
target is an ideal benchmark: given its expected $<10:1$ luminosity
contrast, one should expect a strong, unambiguous Ker-phase signal.

Moreover, ground-based infrared aperture masking interferometry
measurements reported by \citet{2009ApJ...695.1183M} combined with the
astrometry have lead to strong constraints on the orbit of the
companion around the primary. The location of GJ164 B measured from
the Ker-phase analysis of the data can be compared to the orbit
prediction.

Figure \ref{fig:hst_pup} shows the model of the pupil used for this
exercise. The HST pupil exhibits a 30 \% central obscuration as well
as 90$^\circ$ spider arms (actual dimensions were taken from the NIC1 
configuration file in the TinyTim PSF simulation package for HST).
The phase across the pupil is discretized into a 156 elementary
sub-aperture array, whose locations fall on a regular square grid of
step of $1/16^{th}$ of the outer pupil diameter. The phase sample
is assembled into a 155-component ($N-1$) vector $\varphi$.

\begin{figure*}
\plotone{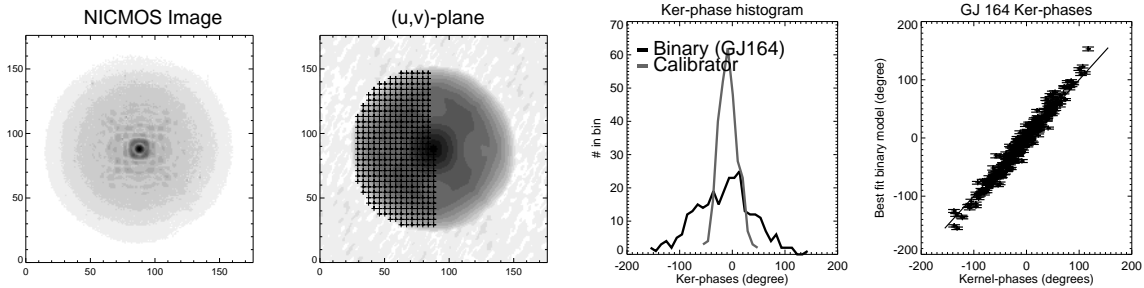}
\caption{
  From left to right: example of a narrow band (F190N) NICMOS1 image
  used for this work, visualized with a non-linear brightness scale;
  The Fourier transform of this image. The 366 sample points for the
  phase in the Fourier domain are overlaid; A comparison of histograms
  of the 288 Ker-phases calculated using the relations identified in
  Section \ref{sec:ker}. By design, the Ker-phases calculated from
  images of a single star are expected to be zero within uncertainty:
  the corresponding histogram (gray curve) confirms this expectation.
  In comparison, the Ker-phase histogram of the binary (dark curve)
  appears significantly larger.
  The same GJ164 Ker-phases plotted against the model of a binary star
  that best fits the data achieve to convince of the presence of a
  companion in the data.
}
\label{fig:data}
\end{figure*}

These 156 pupil phase samples map in the $(u,v)$-plane onto a square
grid of 366 distinct elements \footnote{Note for reference, that a
non-redundant array of 156 sub-apertures would produce exactly 12,090
distinct $(u,v)$ points.}. The resulting $(u,v)$-sampling is
illustrated in Fig. \ref{fig:data}. For this analysis, $\mathbf{A}$
(cf. Eq. \ref{eq:transfer}) is therefore a $155 \times 366$
rectangular matrix, whose SVD reveals that 78 singular values are
non-zero, leaving $366-78=288$ Ker-phase relations.

The GJ164 data consists of a total of 80 frames, acquired at average
Julian Date 2453049.3 (February 14, 2004 UT). Each image is a
non-saturated 32 second exposure, and the target was acquired in a
total of 10 distinct dither positions. Note that this dataset does not
include images on a point-source and therefore, the Ker-phases
calculated from this dataset are non-calibrated. 
Images corresponding to one dither position were simply coadded
forming a final total of ten 250-second exposure images, and assembled
into a datacube. The SAO 179809 data consists of four distinct
20-second exposure frames assembled into a separate datacube.

For both datacubes, the images were then centered with sub-pixel
accuracy and windowed by a super-Gaussian function as described by
\citet{2008ApJ...679..762K} to limit sensitivity to readout noise. The
window size is about $25 \lambda/D$ in diameter, which is significantly
larger than the field of view in which this technique is relevant.

After this preparatory stage, the images are simply
Fourier-transformed (cf. second panel of Fig. \ref{fig:data}), and the
signal $\mathbf{R} \cdot \Phi$ is directly measured for each of the
366 $(u,v)$ points by sampling the imaginary part of the local complex
visibility.
The uncertainty associated to the measurement of each phase is
estimated from the dispersion of the signal in the direct neighborhood
of the $(u,v)$ point.

The $(u,v)$ signals are then assembled into Ker-phases using the
relations gathered in the rows of $\mathbf{K}$ and uncertainties are
propagated.
The procedure is repeated for each of the frames within each
datacube. The final retained series of 288 Ker-phases is the weighted
average for all frames.

Because the Ker-phase relations are designed to produce quantities
independent from pupil phase errors, a point source is expected to
exhibit zero signal within uncertainty. Despite the small number
statistics (four frames acquired on SAO 179809), the Ker-phase of the
calibrator do average to zero (with a $19.7^{\circ}$ standard
deviation), while the binary exhibits a large signal amplitude
($>100^\circ$) in comparison with the uncertainty of individual
Ker-phases ($\sim 2^\circ$). The third panel of Fig. \ref{fig:data}
compares the Ker-phase histograms of both datasets.

To further investigate the GJ 164 data, a parametric model of the
$(u,v)$-plane phase $\Phi_O$ for a binary star is needed. The
parameters are: the angular separation, the position angle of the 
secondary relative to the primary and the luminosity contrast
ratio. The model phase $\Phi_O$ is then multiplied by the diagonal
matrix $\mathbf{R}$, and finally, transformed into model Ker-phases
using the relations established during the SVD.

\begin{figure}
\center{\resizebox*{0.5\columnwidth}{!}{
    \includegraphics{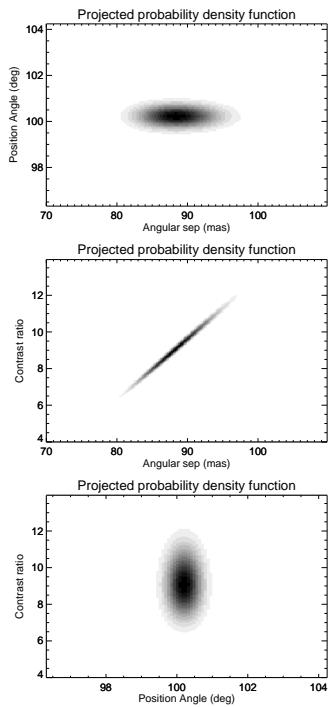}}}
\caption{
  To determine confidence intervals for the parameters of the binary,
  a likelihood analysis comparable to the one presented by
  for closure-phase was performed. These
  panels show the three projections of the likelihood function in the
  region of best fit. Except for the expected correlation between
  angular separation and contrast ratio for a detection within 1
  $\lambda/D$, the solution is unambiguous and well constrained,
  demonstrating the elegance of the Ker-phase approach.
}
\label{fig:likely}
\end{figure}

The agreement between the data and the model is very good (cf. panel 4
of Fig. \ref{fig:data}),
considering the large number of measurements (288) adjuted by only
three parameters. The uncertainties on the Ker-phases, determined from
the scatter of the data however lead to a best fit reduced $\chi^2$
larger than one. A global error term ($10^{\circ}$) is then added in
quadrature to the uncertainty to account for a systematic error in the
non-calibrated Ker-phase and produce a final reduced $\chi^2=1$.

One can then proceed with determining the uncertainty on the
parameters of the model fit, by close examination of the likelihood
function, very much like what is described in
\citep{2009ApJ...695.1183M}.
The three panels of Fig. \ref{fig:likely} show the evolution of this
function in the parameter space region near the best solution.
Just like with closure-phase data, at angular separations less than 1
$\lambda/D$, contrast and separation appear to be correlated.

The uncertainty on each parameter of the model-fit is determined after
marginalization of the likelihood function over the other two
parameters. Despite the noted correlation, the constraint on the
parameters appears satisfactory, and the best fit (cf. Table
\ref{tbl:params}) lies well within one $\sigma$ of the position
predicted from the orbital parameters determined from NRM
interferometry from the ground.
It also matches the location reported by \citet{2004ApJ...617.1323P},
after substraction of a simulated PSF from the same data, only with a
constraint on the position angle improved by a factor of 10.

\begin{deluxetable}{lr@{ $\pm$ }lc}
\tablecolumns{4}
\tablewidth{0pc}
\tablecaption{KER-PHASE DETECTION OF GJ164B IN NICMOS DATA COMPARED
TO PREDICTION FROM ORBITAL PARAMETERS}
\tablehead{
  \colhead{Parameter} & 
  \multicolumn{2}{c}{Ker-phase fit} & \colhead{Prediction}
}
\startdata
Sep. (mas)     &   88.5  & 3.6  & 88.2 \\ 
P.A. (degrees) &  100.6  & 0.3  & 100.4 \\
Contrast       &    9.1  & 1.2  &
\enddata
\label{tbl:params}
\end{deluxetable}

From its (H-K) color index, \citet{2009ApJ...695.1183M} were able to
conclude that GJ 164 B is of spectral type later than M8.5, while the
primary is well characterized as a M4.5 dwarf.
One of the most prominent spectral features for M dwarfs is the broad
absorption band of water at 1.8 $\mu m$, getting deeper with later
types \citep{1994MNRAS.267..413J, 2001ApJ...548..908L}.
The $\sim 5:1$ contrast ratio quoted in the NRM paper was determined
over the full K$_s$ filter (bandwidth 2.0-2.3 $\mu m$). A careful
examination of the spectral sequence by \citep{1994MNRAS.267..413J}
reveals that for this combination of spectral types, the luminosity of
GJ 164B relative to GJ164A seen in the NICMOS F190N filter is expected
to drop by 30 to 40 \% due to the water absorbtion band. The $9:1$
contrast determined from the Ker-phase model (cf Table
\ref{tbl:params}) in this narrow filter reflects this evolution.
The analysis of this GJ164 data demonstrates the validity of the
Ker-phase approach, by positively detecting a companion whose
existence was known beforehand. This $<10:1$ contrast detection was 
however expected to be easy, despite the small angular separation of
the detection (0.6 $\lambda/D$).

\begin{figure}
\plotone{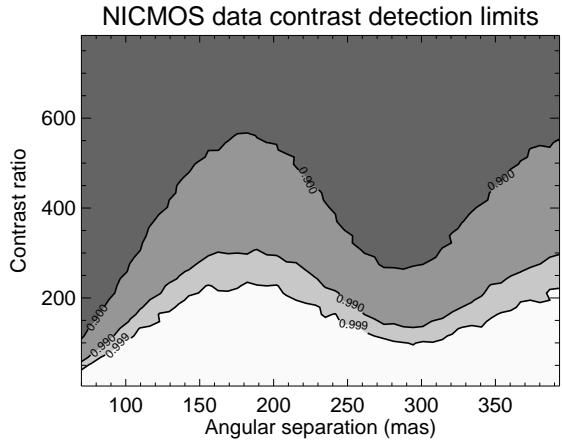}
\caption{Level of confidence in the detection of a companion from the
  analysis of the HST/NICMOS data with the Ker-phase algorithm.
  A darker color indicates a region of lower confidence level. Three
  levels are highlighted: the 90\%, 99\% and 99.9\% confidence levels.
  At angular separation 0.5 $\lambda/D$ (i.e. 80 mas at $\lambda=1.9
  \mu m$), a contrast limit better than 50:1 is possible at the 99\%
  confidence level.
}
\label{fig:sensi}
\end{figure}

Typical NICMOS1 datasets on a given target usually consist of
four frames only. The SAO 179809 dataset is then a representative
example and the statistics of its Ker-phase ($\sigma=19.7^{\circ}$)
can be used in a Monte-Carlo simulation to determine contrast
detection limits.

Because the sampling of the $(u,v)$-plane exhibits no gap, the
sensitivity does not depend on the the position angle relative to the
central star. One can however expect it to be a function of angular
separation. A total of 10,000 simulations were performed per
point in the angular separation/contrast plane to produce the
sensitivity map displayed in Fig. \ref{fig:sensi}. The map highlights
the 90, 99 and 99.9 \% confidence level detection thresholds.

The technique looks promising: for such a dataset, at 0.5
$\lambda/D$, a 50:1 contrast detection appears possible at the 99
\% confidence level. The sensitivity increases and peaks at 180 mas,
which unsurprisingly corresponds to the location of the first zero of
the diffraction for the centrally obstructed telescope (about 1.1
$\lambda/D$), and reaches $\sim$200:1.

\section{Conclusion}
\label{sec:conclu}

Classical closure-phase appears to be one special case of a wider
family of observable quantities that are immune to phase noise and
non-common path errors.
In the high Strehl regime, it was demonstrated that closure-phase like
quantities, called Ker-phases, can be extracted from focal plane
images, and provide high quality ``interferometric grade'' information
on a source, even when the pupil is redundant.
The Ker-phase technique was successfully applied to a series of
archive NICMOS images, clearly detecting a 10:1 contrast companion at
a separation of $0.5 \lambda/D$. Non-calibrated Ker-phase appears
sensitive to the presence of 200:1 contrast companion at angular
separation $1 \lambda/D$. Re-analysis of other comparable NICMOS
datasets with this technique might very well lead to the detection of
previously undetected objects in the direct neighborhood of nearby
stars.

Unlike closure-phases, which are extremely robust to large wavefront
errors, the use of Ker-phases is however for now restricted to the
high-Strehl regime, and will only become relevant to ground based
observations, when extreme AO systems become available.
There is nevertheless hope to be able to extend the application of
Ker-phases to not-so-well corrected AO images, using additional
differential techniques. One possibility, consists in using integral
field spectroscopy, to follow in the Fourier plane, the evolution of
the complex visibilities as a function of wavelength. With enough
resolution and spectral coverage, this indeed allows to identify the
phasors contributing to the power contained at one spatial frequency.

The singular value decomposition of the transfer matrix used to
create Ker-phase relations can also be used to produce a pseudo
inverse to the matrix, and in some cases, allows to inverse the
relation linking the $(u,v)$ phases to the pupil phases. This
means that under certain conditions, a single monochromatic focal
plane image can also be useful for wavefront sensing purposes. This is
particularly interesting since the measurement is happening at the
level of the final science detector, which therefore allows to
calibrate non-common path errors.
The application of the formalism to wavefront sensing will be the
object of a future publication.

\acknowledgements{The author thanks Michael J. Ireland and Olivier
  Guyon for the useful discussions of the ideas presented in this
  work.}

\end{document}